\documentclass[conference]{IEEEtran}

\usepackage[left=0.673in,right=0.660in,top=0.75in,bottom=1.04in]{geometry}

\usepackage[utf8]{inputenc} 
\usepackage[T1]{fontenc}
\usepackage{url}
\usepackage{ifthen}
\usepackage{cite}
\usepackage[cmex10]{amsmath} 

\usepackage{amssymb}
\usepackage{amsthm}
\usepackage{cite}
\usepackage{enumitem}

\usepackage{bm}
\usepackage{stackengine} 
\newcommand\oast{\stackMath\mathbin{\stackinset{c}{0ex}{c}{0ex}{\ast}{\bigcirc}}}

\usepackage[nofillcomment]{algorithm2e}
\RestyleAlgo{ruled}

\newcommand{\myalgorithm}[1]{%
	\begingroup
	\removelatexerror
	\begin{algorithm*}[H]
		#1
	\end{algorithm*}
\endgroup}

\newtheorem{Lemma}{Lemma}
\newtheorem{Theorem}{Theorem}

\newtheorem{Remark}{Remark}
\newtheorem{Definition}{Definition}
\newtheorem{Assumption}{Assumption}

\newcommand{\RQ}{\mathcal{R}_q}
\newcommand{\alsize}{Q}

\newcommand{\compnoisev}{c_{N_v}}
\newcommand{\compnoiseu}{c_{N_u}}

\newcommand{\encmap}{\mathsf{Encode/Map}}
\newcommand{\demapdec}{\mathsf{Demap/Decode}}

\newcommand{\decomp}{\ensuremath{\mathsf{decomp}}_{q}}
\newcommand{\comp}{\mathsf{comp}_{q}}

\DeclareMathOperator{\erfcx_1}{erfcx_1}
\DeclareMathOperator{\erfc}{erfc}

\usepackage{float}
\usepackage{pgfplots}
\usepackage{pifont}
\definecolor{mycolor1}{rgb}{0.00000,0.44700,0.74100}
\definecolor{mycolor2}{rgb}{0.85000,0.32500,0.09800}
\definecolor{mycolor3}{rgb}{0.4660, 0.6740, 0.1880}

\makeatletter
\newcommand{\removelatexerror}{\let\@latex@error\@gobble}
\makeatother
\pgfplotsset{compat=1.18}

\usetikzlibrary{external}
\begin{document}
\title{Reducing Ciphertext and Key Sizes for MLWE-Based Cryptosystems} 

\author{%
  \IEEEauthorblockN{Georg Maringer}
  \IEEEauthorblockA{Institute for Communications Engineering\\
                    Technical University of Munich\\
                    Munich, Germany\\
                    Email: georg.maringer@tum.de}
                    \and
  \IEEEauthorblockN{Antonia Wachter-Zeh}
  \IEEEauthorblockA{Institute for Communications Engineering\\
                    Technical University of Munich\\
                    Munich, Germany\\
                    Email: antonia.wachter-zeh@tum.de}
}

\maketitle

\begin{abstract}
    The concatenation of encryption and decryption can be interpreted as data transmission over a noisy communication channel. In this work, we use finite blocklength methods (normal approximation and random coding union bound) as well as asymptotics to show that ciphertext and key sizes of the state-of-the-art post-quantum secure key encapsulation mechanism (KEM) Kyber can be reduced without compromising the security of the scheme. We show that in the asymptotic regime, it is possible to reduce the sizes of ciphertexts and secret keys by 25\%   for the parameter set Kyber1024 while keeping the bitrate at 1 as proposed in the original scheme. For a single Kyber encryption block used to share a 256-bit AES key, we furthermore show that reductions in ciphertext size of 39\% and 33\% are possible for Kyber1024 and Kyber512, respectively.
\end{abstract}

\section{Introduction}
\label{sec:introduction}
The security of most currently deployed cryptographic schemes relies on the hardness of the integer factorization problem, e.g., RSA~\cite{rivest1978method}, or on the hardness of solving the discrete logarithm problem. However, in 1999, Shor \cite{shor1999polynomial} proposed an algorithm that runs in polynomial time on a quantum computer. Understanding that the development of more sophisticated quantum computers with a larger amount of qubits causes a significant threat to the security of currently deployed systems, the NIST started a competition \cite{NIST_PQC} for key encapsulation mechanisms (KEMs) and digital signatures that can withstand attacks performed on a quantum computer.

Several algorithms have been proposed, and in fact, regarding key encapsulation, an algorithm named Kyber \cite{Kyber_supp} has already been standardized. The security of this algorithm relies on a hard mathematical problem related to lattices, specifically the so-called Module Learning with Errors (MLWE) problem. Several other algorithms relying on similar lattice-based problems have also been proposed within the competition, e.g., NewHope \cite{alkim2019newhope}, LAC \cite{LAC_supp} and Frodo \cite{bos2016frodo}.
Our proposed framework is quite general, and therefore, the same ideas can be applied to all of the aforementioned schemes.

In \cite{lee2019modification}, it has been shown that the concatenation of encryption and decryption in Frodo can be interpreted as a digital communication system. In \cite{maringer2020higher} it has been suggested to view this concatenation as data transmission over a noisy channel and a lower bound on the channel capacity has been established. Furthermore, under the assumption of independent decryption failures and by using BCH code constructions, it has been shown that the number of message bits per ciphertext bit can be significantly enhanced compared to the originally proposed schemes. Our BCH code construction was later improved by adding Gray Coding and vector quantization \cite{bocharova2022coding}.

While this may be useful for certain applications, especially when the scheme is used directly for data transmission rather than for key exchange, a perhaps even more relevant task for KEMs is to decrease the ciphertext and key sizes for a fixed amount of message bits per ciphertext block. This is particularly relevant in the most common scenario where a KEM is used to share the $128$-bit or $256$-bit secret key of a symmetric block cipher, e.g., AES \cite{daemen1998block}. This symmetric cipher can then be used to encrypt the communication as its complexity is typically much smaller than the one of public-key algorithms.

This work proposes a methodology that decreases the ciphertext size of Learning with Errors (LWE)-based encryption schemes without compromising their security. We present asymptotic and finite length results utilizing the normal approximation and the RCU bound \cite{polyanskiy2010channel} showing that ciphertext and key sizes can be significantly reduced for Kyber.

In Section~\ref{sec:preliminaries}, we introduce the notation used throughout this work, present some basics about the LWE problem, and introduce the necessary communication theoretic background. Section~\ref{sec:mlwe_cryptosystems} shows the key generation, encryption, and decryption of a typical MLWE-based encryption scheme. In Section~\ref{sec:lowerbound_capacity}, we show methods to decrease ciphertext and key sizes without compromising the scheme's security and present asymptotic achievability results. Section~\ref{sec:finite_length} deals with the problem of reducing ciphertext and key sizes for a single Kyber ciphertext block typically used to share a 256-bit private key over a public channel. Section~\ref{sec:conclusion} summarizes the results and concludes the paper.

\vspace{-0.1em}
\section{Preliminaries}\label{sec:preliminaries}
We define the polynomial ring $\mathcal{R}_q:=\mathbb{Z}_q[x]/(x^n+1)$.
We denote polynomials by lowercase letters, i.e., the polynomial $a$ and its $i$-th coefficient by $a_i$. Vectors containing polynomials are denoted by bold lowercase letters, e.g., the vector $\bm{v}$, while its $i$-th component is denoted by $v_i$. Matrices are specified by bold uppercase letters, e.g., the matrix $\bm{A}$, while $a_{ij}$ denotes its element in row $i$ and column $j$. In this work, we declare the binomial distribution by $\mathcal{B}(i,n,p)$ with $i$ specifying the number of successful trials, $n$ denoting the total number of trials, and with $p$ the success probability. 

We denote the sampling of an element $x$ from a distribution $\chi$ by $x \xleftarrow[]{\text{\$}} \chi$ and the uniform sampling of an element from a set $\mathcal{S}$ by $x \xleftarrow[]{\text{\$}} \mathcal{S}$. 
The magnitude of an element in $\mathbb{Z}_q$ is defined by the magnitude of the element's representation in $[-q/2,q/2]$. We declare the rounding operator $\lceil . \rfloor$, where ties are rounded up, i.e. $\lceil x.5 \rfloor = x+1$.

To denote the convolution of two probability mass functions, we use the $*$-operator, e.g., $P_X*P_Y$ denotes the convolution of the probability mass functions $P_X$ and $P_Y$. The $n$-fold convolution of a probability mass function $P_X$ with itself is declared by $\oast_n(P_X) := P_X * P_X * \dots * P_X$, in particular $\oast_1(P_X) = P_X * P_X$ and $\oast_0(P_X) = P_X$.

The following distribution is used in many LWE-/MLWE-based encryption schemes.
\begin{Definition}
	The centered binomial distribution $\chi_k$ with parameter $k$ is defined by $\chi_k:=\mathcal{B}(x+k/2,k,1/2)$, where $x \in \{-k/2,-k/2+1,\ldots,k/2\}$.
\end{Definition}

The security of many encryption and signature schemes is based on hard problems on lattices. For an overview on that topic, we refer to \cite{hoffstein2008introduction, peikert2016decade}.
 
The hardness of Kyber is based on the following MLWE problem.
\begin{Definition}
	The decisional version of the MLWE problem is to distinguish a set of samples of the form
	\begin{equation*}
		(\bm{a}_i,b_i=\bm{a}_i\bm{s}+e_i),\quad i=1,\ldots,m \enspace ,
	\end{equation*}
	where the vectors $\bm{a}_i$ are sampled from the uniform distribution on $\RQ^l$, $\bm{s}$ is sampled from $\chi_k(\RQ^l)$ and the $e_i$ are sampled from $\chi_k(\RQ)$.
\end{Definition}

\begin{Definition}
    We define the bitrate $R$ of a code of length $n'$, also called the rate in bits, by
    \begin{equation}
        R:=\frac{\log_2(M)}{n'} \enspace ,
    \end{equation}
    where $M$ denotes the cardinality of the code.
\end{Definition}
\begin{Remark}
    Notice that we used the letter $n'$ to denote the block length rather than $n$, which specifies the number of coefficients within the polynomials used in this work.
\end{Remark}

\section{MLWE-based Cryptosystems}\label{sec:mlwe_cryptosystems}

\subsection{Ciphertext compression and decompression}
Before describing MLWE-based cryptosystems, we will describe ciphertext compression and decompression as they are components of Kyber's encryption and decryption functions.

Ciphertext compression and decompression in MLWE-based schemes like Kyber have the purpose to reduce the ciphertext text. Note that they are not performed to make the system secure in comparison to Learning with Rounding schemes like Saber \cite{d2018saber}. 

Let $z$ denote the input of the compression function and $z'$ its output. The parameter $d_c$ determines the number of bits to which the input is compressed. This is achieved by the following definition of the compression function:
\begin{equation*}
z' = \comp(z,d_c) = \left \lceil \frac{z \cdot 2^{d_c}}{q} \right \rfloor \mod 2^{d_c}.
\end{equation*}
To decompress $z'$ to $z''$ we define the decompression function by
\begin{equation*}
z'' = \decomp(z',d_c) = \left \lceil \frac{z' \cdot q}{2^{d_c}} \right \rfloor \enspace .
\end{equation*}
To use these functions, its inputs ($z$ and $z'$) are considered in their representation in $\{0,\ldots q-1\}$.
The concatenation of $\comp$ and $\decomp$ is lossy, and therefore, decryption is more likely to become incorrect if the ciphertext compression is enhanced in MLWE-based schemes. 

\subsection{Description of MLWE-based public key encryption schemes}
A public-key encryption scheme consists of three algorithms: Key Generation, Encryption, and Decryption. They are presented in Algorithms~\ref{algorithm:key_gen_MLWE}, \ref{algorithm:encryption_MLWE} and \ref{algorithm:decryption_MLWE}.

\begin{figure}[t]
	\begin{center}
		\begin{minipage}{\columnwidth}
			\myalgorithm{
				\small
				\KwIn{$n,q,k_1,l$}
				$\boldsymbol{A} \xleftarrow[]{\text{\$}} \RQ^{l \times l}$ \\
				$\boldsymbol{s},\boldsymbol{e} \xleftarrow[]{\text{\$}} \chi_{k_1}(\RQ^l)$\\
				$\boldsymbol{b} \leftarrow \boldsymbol{A}\boldsymbol{s}+\boldsymbol{e}$\\
				\KwResult{$pk=(\bm{A}, \bm{b})$, $sk = \boldsymbol{s}$}
				\caption{Key Generation}
				\label{algorithm:key_gen_MLWE}
			}
			\myalgorithm{
				\small
				\KwIn{$pk=(\bm{A},\bm{b})$, $m \in \mathcal{M}$, $(n,q,k_1,k_2,l)$, $d_u,d_v$}
				$\boldsymbol{s'} \xleftarrow[]{\text{\$}} \chi_{k_1}(\RQ^l)$ \\
                $\boldsymbol{e'} \xleftarrow[]{\text{\$}} \chi_{k_2}(\RQ^l)$ \\
				$e'' \xleftarrow[]{\text{\$}} \chi_{k_2}(\RQ)$\\
				$\boldsymbol{u} \leftarrow \bm{A}^T\bm{s'} + \boldsymbol{e'}$ \\
				$v \leftarrow \bm{b}^T \bm{s'} + e'' + \encmap(m)$\\
				$\bm{u'} \leftarrow \comp(\bm{u},d_u)$\\
				$v' \leftarrow \comp(v,d_v)$\\
				\KwResult{$c= (\bm{u'}, v')$} 
				\caption{Encryption}
				\label{algorithm:encryption_MLWE}
			}
			\myalgorithm{
				\small
				\KwIn{$c= (\bm{u'}, v')$, $sk=\boldsymbol{s}$, $(n,q,k_1,k_2,l)$, $d_u,d_v$}
				$\bm{u''} \leftarrow \decomp(\bm{u'},d_u)$\\
				$v'' \leftarrow \decomp(v',d_v)$\\
				$\hat{m} \leftarrow \demapdec(v'' - \bm{s}^T \bm{u''})$\\
				\KwResult{$\hat{m}$}
				\caption{Decryption}
				\label{algorithm:decryption_MLWE}
			}
			\vspace{-2em}
		\end{minipage}
	\end{center}
\end{figure}

The structure of LWE, Ring Learning with Errors (RLWE), and MLWE-based schemes is very similar, e.g., to switch from an MLWE to an RLWE-based scheme, one simply chooses the parameter $l=1$ in all of the aforementioned algorithms.

The key generation (Algorithm~\ref{algorithm:key_gen_MLWE}) uniformly samples an $l\times l$ matrix of elements in $\mathcal{R}_q:=\mathbb{Z}_q[x]/(x^n+1)$. Then two vectors of elements $\bm{s},\bm{e} \in \RQ$ are sampled from the centered binomial distribution $\chi_k$ and the second component of the public key $\bm{b}=\bm{A}\bm{s}+\bm{e}$ is computed. The public key is the pair $(\bm{A},\bm{b})$ and the private key is $\bm{s}$.
\begin{Remark}
	In practice, the matrix $\bm{A}$ is generated using a pseudo-random number generator (PRNG), which uses a true random seed as its input. Since the PRNG is a deterministic device, knowledge of the seed is sufficient to generate the matrix $\bm{A}$, and it is common to share the pair $(seed,\bm{b})$ as the user's public key to save communication bandwidth.
\end{Remark}

The encryption function (Algorithm~\ref{algorithm:encryption_MLWE}) obtains the public key and a secret message $m$ out of the message space $\mathcal{M}$ as input. The algorithm consists of elementary arithmetic steps except for the encoding and mapping of the message $\encmap(m)$ into $\RQ$ and the ciphertext compression $\comp$.
$\encmap(.)$ combines the encoding of the message and the mapping onto the channel input. The mapper maps the $\alsize$-ary encoder's output symbol $x\in \{0,\ldots,\alsize-1\}$ either onto $\lfloor x\cdot q/\alsize \rfloor$ or onto $\lceil x \cdot q/\alsize \rceil$. Basically, the mapper tries to achieve as equal spacing as possible in the set $\{0,\ldots,q-1\}$. 

\begin{figure}[t]
    \vspace{0.5em}
	\begin{center}
		\scalebox{0.87}{
			\includegraphics{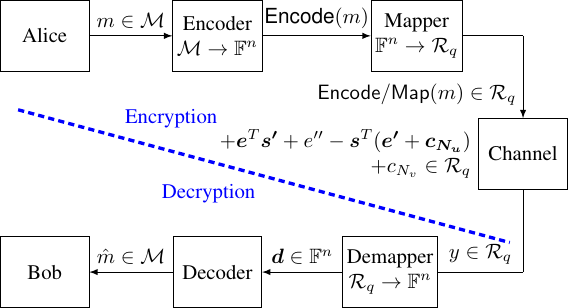}
			}
	\end{center}
	\caption{MLWE Channel}
	\label{fig:RLWE_channel}
    \vspace{-1em}
\end{figure}

The decryption function (Algorithm~\ref{algorithm:decryption_MLWE}) obtains the ciphertext and makes use of the private key to recover the secret message $m$. Basically, throughout the decryption, both ciphertext components are first decompressed, and then demapping and decoding are applied on $v''-\bm{s}^T\bm{u''}$. Having a closer look at this expression one obtains that
\begin{align}\label{eq:additive_noise_mlwe}
	v''-\bm{s}^T\bm{u''} = &\ \encmap(m) + \bm{e}^T\bm{s'} \nonumber\\ & - \bm{s}^T (\bm{e'} + \bm{\compnoiseu}) + e'' + \compnoisev \enspace .
\end{align}
Equation~\eqref{eq:additive_noise_mlwe} motivates why the concatenation of encryption and decryption in MLWE-based cryptosystems can be interpreted as data transmission over a channel with additive noise. The resulting channel is illustrated in Fig.~\ref{fig:RLWE_channel}.

The standard way to decode $m$ is to round symbolwise to the nearest symbol that the mapper may have outputted. Then the mapping function is reversed, and the resulting vector is forwarded as an input to the decoder, which outputs an estimate for the message $\hat{m}$. This procedure gives the correct result with high probability because all components of the noise are small in magnitude for appropriately chosen parameters $k,d_u$, and $d_v$. Therefore, the magnitude of the entire channel noise is small with high probability and the rounding gives a reasonable estimate of the correct symbols.

\begin{Lemma}[Lemma 3 \cite{maringer2022information}]\label{lem:additive_noise}
	Let the distribution of the product of two elements be sampled from the centered binomial distribution $\chi_k$ be denoted as $\xi_k$. Let the distribution of one coefficient in $\bm{s}^T(\bm{e'}+\bm{\compnoiseu})$ be labelled as $\eta_k$ and the distribution of $\compnoisev$ as $\rho_v$. Then, the probability mass function of the additive noise within the MLWE channel can be computed as
	\begin{equation*}
		\psi = \oast_{l-1}(\oast_{n-1}(\xi_k))  * \eta_k * \chi_k * \rho_v \enspace .
	\end{equation*}
\end{Lemma} 

\begin{Theorem}[Theorem 4 \cite{maringer2022information}]\label{th:asympt_lower}
    Let $n$ denote the length of one MLWE block (length of one polynomial in $\RQ$), and let $\psi$ be the distribution of the additive noise of the MLWE channel specified in Lemma~\ref{lem:additive_noise}. For the capacity of the MLWE channel it holds that
    \begin{align}\label{eq:lower_bound_asympt}
	{C}_{\mathsf{MLWE}} &:= \max_{P_{X^n}} I(X^n;Y^n) \geq {C}_{\mathsf{MLWE}}^{marg} := n \max_{P_X}           I(X;Y) \nonumber \\
        &\geq n \left ( H\left( \frac{1}{\alsize} \sum_{j=0}^{\alsize - 1} \psi_j \right) - H(\psi) \right) \enspace ,
    \end{align}
    where $\psi_j(x) := \psi(x+\lfloor jq/\alsize \rfloor)$ and $I(X;Y)$ is determined by the marginal distribution $P(Y_1|X_1)=\ldots=P(Y_n|X_n)$.
\end{Theorem}

\section{Reducing Ciphertext and Key sizes and Asymptotical Setting}\label{sec:lowerbound_capacity}
The asymptotic achievability bound for Kyber from \cite{maringer2022information} can be used to find parameters that significantly reduce ciphertext and key sizes in the asymptotic regime when a large amount of ciphertext blocks are concatenated and the messages within all blocks are encoded using an error-correcting code (ECC) of suitable blocklength.

In the following, we show how to modify the parameters of Kyber without compromising the schemes' security while still keeping the bitrate to be at least 1. We decided not to modify the parameters $n$ and $q$ because its relation enables using the number theoretic transform (NTT) for the required polynomial multiplications. Similarly to the FFT it is possible to perform convolutions in approximately $\mathcal{O}(n\log(n))$ instead of $\mathcal{O}(n^2)$.

\subsection{Decreasing keys and ciphertexts by modifying $l$}
A natural way to decrease both ciphertext and key sizes is to modify the MLWE parameter $l$. However, not changing other parameters accordingly will decrease the security level of the scheme. Our approach to this problem is to increase the parameter $k$, which specifies the centered binomial distribution, to compensate for this effect. This, in turn, increases the Decryption Failure Rate (DFR), which needs to be compensated by choosing an appropriate ECC, as decryption failures provide exploitable information to an attacker \cite{fluhrer2016cryptanalysis}. Furthermore, the standard way of transforming a chosen plaintext secure public key encryption scheme into a chosen ciphertext secure (IND-CCA2) key encapsulation mechanism requires a DFR of approximately $2^{-\text{security level}}$ \cite{hofheinz2017modular}. Using the results presented in \cite{albrecht2015concrete}, it is possible to estimate the hardness of the underlying lattice problems of the algorithm for the modified parameters.
The authors also provide a sage tool that implements their work, which we used to estimate the security of our modified parameters.
The lower bound on the capacity of the resulting MLWE channel gives insight into whether such an ECC exists.

In this section, we try to reach the security level of the parameter set Kyber1024 $(n,q,l,d_u,d_v,k_1,k_2) = (256,3329,4,11,5,2,2)$ after the modification of $l=3$. To compensate for the loss in security, the parameters $k_1,k_2$ of the centered binomial distribution have to be increased accordingly. We obtained that $k_1=k_2=14$ is the smallest value that achieves the desired level of security.

\begin{figure}[t!]
	\centering
	\scalebox{0.9}{
		\includegraphics{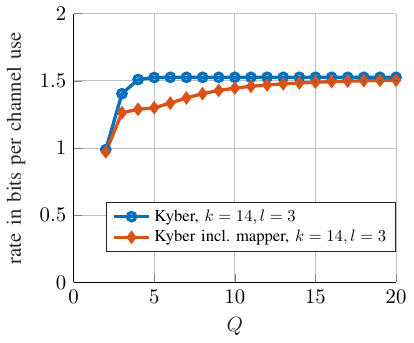}
		}
        \vspace{-1em}
	\caption{Lower bounds on the capacities of the MLWE channel for Kyber with $l=3$ and $k=14$ achieving the security of the Kyber1024 parameter set}
	\label{fig:lower_bounds_Kyber_l}
 \vspace{-1em}
\end{figure}

Fig.~\ref{fig:lower_bounds_Kyber_l} shows that it is possible to achieve bitrates greater than $1$ for this change in parameters if the alphabet size of the encoder is increased appropriately. This parameter set decreases the size of the private key and the ciphertexts by $25\%$, whereas the number of elements in $\bm{A}$ is decreased by about $44\%$. Even though the matrix $\bm{A}$ is generated from a pseudorandom number generator using a seed of fixed length, the amount of storage on the device running the algorithm is significantly reduced.

\subsection{Decreasing the ciphertext size by modifying the compression parameters $d_u$ and $d_v$}
\begin{figure}[t!]
	\centering
	\scalebox{0.9}{
		\includegraphics{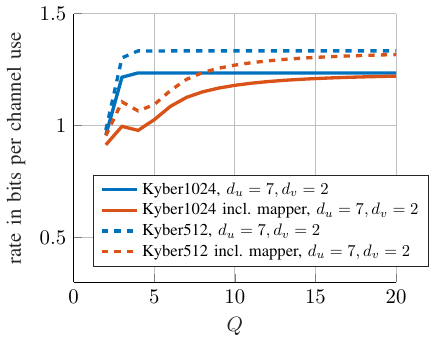}
		}
            \vspace{-1em}
	\caption{Lower bounds on the capacities of the MLWE channels for enhanced ciphertext compression for the Kyber512 and Kyber1024 parameter}
	\label{fig:lower_bounds_Kyber_compression}
    \vspace{-1em}
\end{figure}

In the following, we will examine the parameter sets Kyber512 $(n,q,l,d_u,d_v,k_1,k_2) = (256,3329,2,10,4,2,3)$ and Kyber1024 that achieve the smallest and highest security levels proposed within the NIST PQC standardization.
The results in Fig.~\ref{fig:lower_bounds_Kyber_compression} show that asymptotically, it is possible to significantly enhance the ciphertext compression by using $d_u=7$ and $d_v=2$ for both parameter sets. This reduces the ciphertext by about $39\%$ for Kyber1024 and by about $33\%$ for Kyber512. The difference between the two curves in the figure is that Kyber incl. mapper refers to the case where mapper and demapper are included in the channel. In this case, soft information can only be used during decoding rather than during the combined process of demapping and decoding.

\section{Finite-length Achievability Bounds}\label{sec:finite_length}

In the proof of Theorem~\ref{th:asympt_lower} a memoryless channel having the same conditional distribution over a single channel use (marginalized channel) has been investigated. It has been shown that the capacity of the resulting channel is a lower bound on the capacity of the MLWE channel and that the right-hand side of equation~\eqref{eq:lower_bound_asympt} lower bounds this capacity. In~\cite{d2019impact} it has been shown that assuming independent coefficient failures can lead to an overestimation of the security level due to a higher decryption failure rate of the scheme. The two results do not contradict each other as the decoder is matched to a memoryless channel, which is not present in practice (mismatched decoding).

\begin{Definition}
    The smallest achievable block error probability for a channel $\mathcal{C}$ and any code of cardinality $M$ and blocklength $n'$ under maximum likelihood decoding is $\varepsilon^{\mathcal{C}}_{min}(n',M)$.
\end{Definition}
\begin{Assumption}\label{ass:indep}
    We assume that for the MLWE channel $\mathcal{C}_{\mathsf{MLWE}}$ it holds
    that the lowest achievable block error probability $\varepsilon^{{\mathcal{C}}_{\mathsf{MLWE}}}_{min} (n',M)$ for fixed codebook size $M$ and blocklength $n'$ is lower bounded by the cardinality of the largest codebook for the marginalized MLWE channel ${\mathcal{C}}_{\mathsf{MLWE}}^{marg}$ with the same blocklength and codebook cardinality, i.e.,
    \begin{equation*}
        \varepsilon_{min}^{{\mathcal{C}}_{\mathsf{MLWE}}}(n',M) \geq \varepsilon_{min}^{{\mathcal{C}}_{\mathsf{MLWE}}^{marg}}(n',M) \enspace .
    \end{equation*}
    This is the analog in the finite blocklength regime of the first inequality in equation~\eqref{eq:lower_bound_asympt}.
\end{Assumption}

In the following, we motivate why Assumption~\ref{ass:indep} is reasonable. First, in the asymptotic setting, the statement holds and has been proved (see Theorem~\ref{th:asympt_lower}). Furthermore, this assumption is weaker than the frequently used independence assumption of coefficient failures which indeed has been shown not to hold in general. The construction of BCH codes with their usual encoders and decoders typically does not take into account the channel's memory, i.e., the decoder assumes a memoryless channel, and therefore, we are in a mismatched decoding scenario. That means even though for the channel with memory, it may theoretically be possible to achieve better error performance, due to the suboptimal decoding strategy, the system can have a larger block error probability.

If Assumption~\ref{ass:indep} holds, the result for BCH codes assuming independence still gives an upper bound on the achievable error probability for the MLWE channel. This result has to be interpreted with care, though, as this only means that there exists an encoder/decoder pair achieving at least the same error performance for the MLWE channel (including the memory). However, this does not imply that using the same BCH code construction (including the decoding strategy) achieves at least the same block error probability due to the channel mismatch. Constructing a BCH code and estimating the block error probability for the marginalized MLWE channel we obtain for $\alsize=3$ that it is indeed possible to use stronger ciphertext compression by choosing $d_u=9$ and $d_v=3$ for the Kyber512 parameter set, thereby reducing the ciphertext size by 12.5\%. For a more elaborate discussion on creating results using BCH codes for the MLWE channel see~\cite{maringer2022information}.

\begin{figure}[t]
	\centering
	\scalebox{0.9}{
		\includegraphics{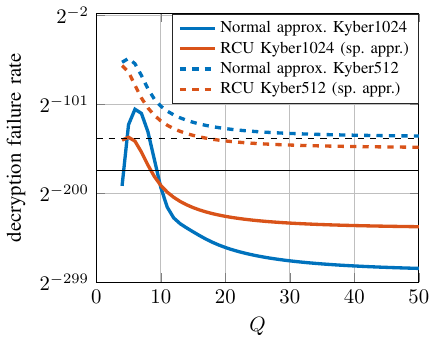}
		}
            \vspace{-1em}
	\caption{RCU bound and normal approximation for the Kyber512 parameter set with $d_u=7,d_v=4$ and for the Kyber1024 parameter set with $d_u=8,d_v=3$, required DFRs for Kyber512 (black, dashed) and Kyber1024 (black)}\label{fig:lower_bounds_Kyber_finite}
\end{figure}
\begin{figure}[t] 
    \centering
    \scalebox{0.9}{
    \includegraphics{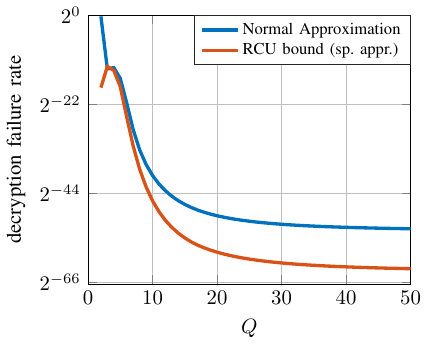}
    }
    \vspace{-1em}
    \caption{RCU bound and normal approximation for Kyber with $l=3$ and $k=14$ achieving the security level of the Kyber1024 parameter set}
    \label{fig:rcu_normal_L3}
    \vspace{-1.1em}
\end{figure}

In this section we present results obtained using the normal approximation and the random coding union bound (RCU bound) \cite{polyanskiy2010channel} to get more precise results for the rather short blocklength regime our schemes operate in.
\begin{Theorem}[Normal approximation]\label{th:normal_approx}
    For a discrete memoryless channel $\mathcal{C}$ with input $X$ and output $Y$ it holds that
    \begin{equation*}
        \log(M^{\mathcal{C}}(n',\varepsilon)) \geq n'I(X;Y) - \sqrt{n'V} Q^{-1}(\varepsilon) + \mathcal{O}(1)\enspace,
    \end{equation*}
    where $M^\mathcal{C}(n',\varepsilon)$ denotes the maximal cardinality of a code of length $n'$ achieving block error probability $\varepsilon$ and $Q^{-1}(.)$ denotes the inverse Q-function. Furthermore, 
    \vspace{-0.4em}
    \begin{equation*}
        V := \sum_{\substack{x \in \mathcal{X}\\ y \in \mathcal{Y}}} P_X(x) P_{Y|X}(y|x) \log^2\left(\frac{P_{Y|X}(y|x)}{P(y)}\right) - I(X;Y)^2 \enspace .
    \end{equation*}
\end{Theorem}

The normal approximation only approximates the maximal cardinality of the code (even though the approximation is rather tight). We next aim at computing an achievability bound, the RCU bound introduced in \cite{polyanskiy2010channel}.
Direct computation of the RCU bound is impractical even for moderately complicated channels. Hence, it has been proposed in \cite{martinez2011saddlepoint,font2018saddlepoint} to use the saddlepoint approximation to approximate the RCU.
\begin{Definition}
	We define Gallager's $E_0(\rho,P_X)$ function to be
	\begin{equation*}
		E_0(\rho,P_X) := -\log_2\left(\sum_{y \in \mathcal{Y}}\sum_{x\in \mathcal{X}} P_X(x) P_{Y|X}(y|x)^{\frac{1}{1+\rho}} \right)^{1+\rho} \enspace .
	\end{equation*}
	
	Further we define $\hat{\rho} := \min(1,\hat{\rho}_0)$        with $\hat{\rho}_0$ as the root of
	\begin{equation*}
		\frac{dE_0(\hat{\rho},P_X)}{d\rho} = R = \frac{1}{n'} \log(M) \enspace ,
	\end{equation*}
	\begin{equation*}
		W := R - \frac{dE_0(\hat{\rho},P_X)}{d\rho},\quad V := - \frac{d^2E_0(\hat{\rho},P_X)}{d^2\rho} \enspace .
	\end{equation*}
\end{Definition}
\begin{Theorem}[\cite{martinez2011saddlepoint,font2018saddlepoint}]\label{th:saddlepoint}
	The RCU bound for a DMC with input $X$ and output $Y$ can be approximated by
	\begin{align}
		\varepsilon \leq e^{n'(\hat{\rho} R - E_0(\hat{\rho},P_X))} \frac{1}{2} \Bigg [\erfcx_1 \left( \hat{\rho} \sqrt{\frac{n'V}{2}}, W\sqrt{\frac{n'}{2V}} \right) \nonumber \\
        + \erfcx_1 \left( (1-\hat{\rho}) \sqrt{\frac{n'V}{2}}, -W\sqrt{\frac{n'}{2V}} \right)\Bigg] \enspace ,
	\end{align}
	where
	\begin{equation*}
		\erfcx_1 (x,y) := \erfc(x-y) \exp(x^2-2xy) \enspace .
	\end{equation*}
\end{Theorem}
For Kyber512 a DFR of $2^{-139}$ and for Kyber1024 a DFR of $2^{-174}$ is necessary according to the specification. Our goal is to share an AES key of 256 Bit. Therefore, we need a bitrate of $1$ if we want to share the key using one MLWE block. Fig.~\ref{fig:lower_bounds_Kyber_finite} shows that for Kyber512, it is possible to use ciphertext compression parameters $(d_u,d_v) = (7,4)$, which corresponds to a reduction in ciphertext size of 25\% and for Kyber1024 a reduction of 28\% for a single MLWE block. Fig.~\ref{fig:rcu_normal_L3} shows that the reduction of the parameter $l$, which is asymptotically possible, cannot be achieved for a single MLWE block.

\section{Conclusion}\label{sec:conclusion}
In this work we have investigated how much we can compress ciphertext and secret key sizes for two parameter sets of Kyber both in the asymptotic as well as in the finite length (one MLWE block) setting. Asymptotically, we have shown that for Kyber1024, the secret key and the ciphertext sizes can be reduced by 25\%. The public matrix $\bm{A}$ shrinks by 44\% in this case. Furthermore, if one is only concerned about the ciphertext sizes a reduction of 39\% and 33\% can be achieved for Kyber1024 and Kyber512, respectively. If one only uses a single Kyber encryption to transmit 256 Bits of plaintext to share an AES key it is possible to reduce the ciphertext sizes by 28\% and 25\%, respectively.

\newpage
\bibliographystyle{IEEEtran}
\bibliography{mybibliography}

\end{document}